\documentclass[10pt]{article}
\usepackage[textwidth=17cm,textheight=22cm]{geometry}
\usepackage{amssymb,amsmath}
\usepackage{graphicx}
\begin{document}
\title{Correlation Distance and Bounds for Mutual Information}

\author{Michael J. W. Hall\footnote{Centre for Quantum Computation and Communication Technology (Australian Research Council), Centre for Quantum Dynamics, Griffith University, Brisbane, QLD 4111, Australia}}
\date{}
\maketitle



\abstract{The correlation distance quantifies the statistical independence of two classical or quantum systems,  via the distance from their joint state to the product of the marginal states. Tight lower bounds are given for the mutual information between pairs of two-valued classical variables and quantum qubits, in terms of the corresponding classical and quantum correlation distances.
These bounds are stronger than the Pinsker inequality (and refinements thereof) for relative entropy.  The classical lower bound may be used to quantify properties of statistical models that violate Bell inequalities.  Entangled qubits can have a lower mutual information than can any two-valued classical variables having the same correlation distance.   The qubit correlation distance also provides a direct entanglement criterion, related to the spin covariance matrix.  Connections of results with classically-correlated quantum states are briefly discussed.}




\section{Introduction}

The relative entropy between two probability distributions has many applications in classical and quantum information theory. A number of these applications, including the conditional limit theorem \cite{cover}, and secure random number generation and communication \cite{hayashi,other}, make use of lower bounds on the relative entropy in terms of a suitable distance between the two distributions.  The best known such bound is the so-called Pinsker inequality \cite{pin}
\begin{equation} \label{pin}
H(P\| Q) := \sum_j P(j) [\log P(j) - \log Q(j)] \geq \frac{1}{2} D(P,Q)^2 \,\log e,
\end{equation}
where $D(P,Q):=\|P-Q\|_1=\sum_j |P(j)-Q(j)|$ is the variational or L1 distance between distributions $P$ and $Q$.  Note that choice of logarithm base is left open throughout this paper, corresponding to a choice of units.  There are a number of such bounds \cite{pin}, all of which easily generalise to the case of quantum probabilities \cite{ohya,rastegin}.  

However, in a number of applications of the Pinsker inequality and its quantum analog, a lower bound is in fact only needed for the special case that the relative entropy quantifies the mutual information between two systems.  Such applications include, for example, secure random number generation  and coding \cite{hayashi, other} (both classical and quantum),  and  quantum de Finnetti theorems \cite{finetti}. Since mutual information is a special case of relative entropy, it follows that it may be possible to find strictly stronger lower bounds for mutual information.  

Surprisingly little attention appears to have been paid to this possiblity of better lower bounds (although upper bounds for mutual information have been investigated \cite{upper}).  The results of preliminary investigations are given here, with explicit tight lower bounds being obtained for pairs of two-valued classical random variables, and for pairs of quantum qubits with maximally-mixed reduced states. 

In the context of mutual information, the corresponding variational distance reduces to the distance between the joint state of the systems and the product of their marginal states, referred to here as the `correlation distance'.  It is shown that both the classical and quantum correlation distances are relevant to quantifying properties of quantum entanglement: the former with respect to the classical resources required to simulate entanglement, and the latter as providing a criterion for qubit entanglement.  In the quantum case, it is also shown that the minimum value of the mutual information can only be achieved by entangled qbuits if the correlation distance is more than $\approx 0.72654$.

The main results are given in the following section.  Lower bounds on classical and quantum mutual information for two-level systems are derived in sections 3 and 5, and an entanglement criterion for qubits in terms of the quantum correlation distance is obtained in Section 4.  Connections with classically-correlated quantum states are briefly discussed in section 6, and conclusions presented in section 7.

\section{Definitions and Main Results}

For two classical random variables $A$ and $B$, with joint probability distribution $P_{AB}(a,b)$ and marginal distributions $P_A(a)$ and $P_B(b)$, the Shannon mutual information and the classical correlation distance are defined respectively by
\begin{eqnarray*}  
I(P_{AB}) &:=& H(P_{AB}\| P_A P_B) = H(P_A) + H(P_B) -H(P_{AB}) ,\\
C(P_{AB}) &:=& \| P_{AB}-P_AP_B\|_1 = \sum_{a,b} \left| P_{AB}(a,b) - P_A(a) P_B(b)\right|,
\end{eqnarray*}
where $H(P):=-\sum_jP(j)\log P(j)$ denotes the Shannon entropy of distribution $P$. The term `correlation distance' is used for $C(P_{AB})$, since it inherits all the properties of a distance from the more general variational distance, and clearly vanishes for uncorrelated $A$ and $B$.  

For two quantum systems $A$ and $B$ described by density operator $\rho_{AB}$ and reduced density operators $\rho_A$ and $\rho_B$, the corresponding quantum mutual information and quantum correlation distance are analogously defined by
\begin{eqnarray*}  
I(\rho_{AB}) &:=& S(\rho_A) + S(\rho_B) -S(\rho_{AB}), \\
C(\rho_{AB}) &:=& \| \rho_{AB} - \rho_A\otimes \rho_B\|_1 = {\rm tr}|\rho_{AB} - \rho_A\otimes \rho_B| ,
\end{eqnarray*}  
where $S(\rho):=-{\rm tr}[\rho \log \rho]$ denotes the von Neumann entropy of density operator $\rho$. 

In both the classical and quantum cases, one has the lower bound 
\begin{equation} \label{pin2} 
I\geq \frac{1}{2}\, C^2 \,\log e
\end{equation} 
for mutual information, as a direct consequence of the  Pinsker inequality (\ref{pin})  for classical relative entropies \cite{pin,ohya,rastegin}.  However, better bounds for mutual information can be obtained, which are stronger than any general inequality for relative entropy and variational distance.  

For example, for two-valued classical random variables $A$ and $B$ one has the tight lower bound
\begin{equation} \label{c2}
I(P_{AB}) \geq \log 2 -  H\left(\frac{1+C(P_{AB})}{2},\frac{1-C(P_{AB})}{2}\right)  
\end{equation}
for classical mutual information.  This inequality has been previously stated without proof in Ref.~\cite{relaxed}, where it was used to bound the shared information required to classically simulate entangled quantum systems. It is proved in section 3 below.

In contrast to Pinsker-type inequalities such as Eq.~(\ref{pin2}), the quantum generalisation of Eq.~(\ref{c2})  is not straightforward.  
In particular, note for a two-qubit system that one cannot simply replace $P_{AB}$  by $\rho_{AB}$ in Eq.~(\ref{c2}), as the right hand side would be undefined for $C(\rho_{AB})>1$ -- which can occur if the qubits are entangled.  Indeed, as shown in section 4, $C(\rho_{AB})>1$ is a sufficient condition for the entanglement of two qubits, as is the stronger condition
\begin{equation} \label{entang}
C(\rho_{AB}) > 2\sqrt{(1-{\rm tr}[\rho_A^2])\,(1-{\rm tr}[\rho_B^2])} .
\end{equation}
An explicit expression for the quantum correlation distance for two qubits, in terms of the spin covariance matrix, is also given in section 4. 

It is shown in section 5 that the quantum equivalent of Eq.~(\ref{c2}), i.e., a tight lower bound for the quantum mutual information shared by two qubits, is 
\begin{equation} \label{q2}
I(\rho_{AB}) \geq \left\{ \begin{array}{ll}
\log 2 -   H\left(\frac{1+C(\rho_{AB})}{2},\frac{1-C(\rho_{AB})}{2}\right)  ,&C(\rho_{AB})\leq C_0,\\
\log 4 - H\left(\frac{1}{4}+\frac{C(\rho_{AB})}{2},\frac{1}{4}-\frac{C(\rho_{AB})}{6},\frac{1}{4}-\frac{C(\rho_{AB})}{6},\frac{1}{4}-\frac{C(\rho_{AB})}{6}\right),&C(\rho_{AB})> C_0,
\end{array} \right. 
\end{equation}
when the reduced density operators are maximally mixed, where $C_0\approx  0.72654$.  For $C(\rho_{AB})>C_0$ this lower bound can only be achieved by entangled states, and cannot be achieved by any classical distribution $P_{AB}$ having the same correlation distance.  It is also shown that, for $C(\rho_{AB})> C_0$, the bound is also tight if only one of the reduced states is maximally mixed.  Support is given for the conjecture that the bound in Eq.~(\ref{q2}) in fact holds for all two-qubit states.  

In section 6 the natural role of `classically-correlated' quantum states, in comparing classical and quantum correlations, is briefly discussed.  Such states have the general form
$\rho_{AB} = \sum_{j,k} P(j,k) |j,k\rangle\langle j,k|$  \cite{horo},
where $P(j,k)$ is a classical joint probability distribution and $\{|j\rangle\}$ and $\{|k\rangle\}$ are orthonormal basis sets for the two quantum systems.  The lower bound in Eq.~(\ref{q2}) can be saturated by a classically-correlated state if and only if $C\leq C_0$.

\section{Tight Lower Bound for Classical Mutual Information}

\subsection{Derivation of Bound}

The tight lower bound in Eq.~(\ref{c2}) is derived here.  The bound is plotted in Figure 1 below [top curve].  Also plotted for comparison are the Pinsker lower bound in Eq.~(\ref{pin2}) [bottom curve], and the lower bound following from the best possible generic inequality for relative entropy and variational distance, given in parametric form in Ref.~\cite{pin} [intermediate curve].

\begin{figure}[h]
\caption{Lower bounds for the classical mutual information between two-valued variables.}
	\centering
		\includegraphics{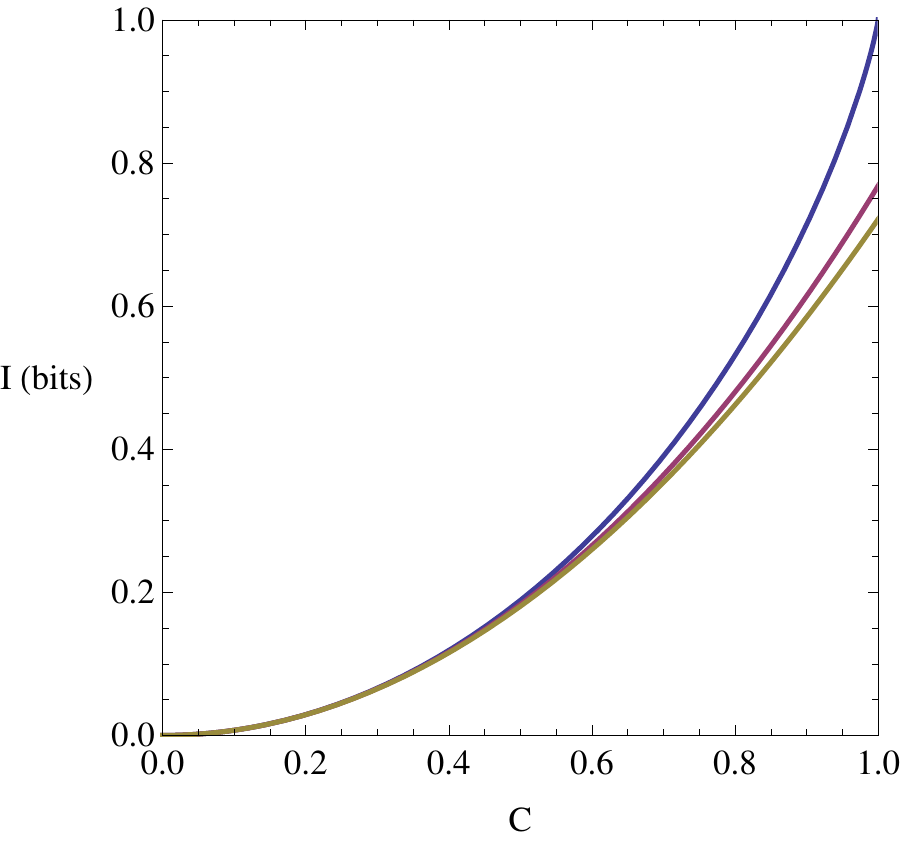}
	\label{fig:pinskerfig1}
\end{figure}

To derive the bound  in Eq.~(\ref{c2}), it is convenient to label the two possible values of $A$ and $B$ by $\pm 1$.  Defining $R(a,b):= 4[P_{AB}(a,b)-P_A(a)P_B(b)]$, it follows by summing over each of $a$ and $b$ that $R(a,b)=abr$ for some number $r$, and hence that $C(P_{AB})=|r|$.  Further, writing $P_A(a)=(1+ax)/2$ and $P_B(b)=(1+by)/2$, for suitable $x,y\in[-1,1]$, the positivity condition $P_{AB}(a,b)\geq 0$ is equivalent to 
\begin{equation} \label{pos}
|x+y| -1 \leq r+xy \leq 1  -|x-y| .
\end{equation}
Now, Eq.~(\ref{c2}) is equivalent to 
\begin{equation} \label{fr} 
f(r):= I(P_{AB}) - \log 2 + H\left( \frac{1+r}{2},\frac{1-r}{2}\right) \geq 0. 
\end{equation}
It is easy to check that this inequality is always saturated for the case of maximally-random marginals, i.e, when $x=y=0$. In all other cases, the inequality may be proved by showing that $f(r)$ has a unique global minimum value of 0 at $r=0$.

In particular, note first that $f(0)=0$ (one has $P_{AB}=P_AP_B$ in this case, so that the mutual information vanishes).  Further, using 
$P_{AB}(a,b) = [(1+ax)(1+by)+abr]/4$, one easily calculates that, using logarithm base $e$ for convenience,
\[ 
f'(r) = \frac{1}{4} \sum_{a,b} ab \log P_{AB}(a,b) -\frac{1}{2} \sum_a a\log \frac{1+ar}{2}=\frac{1}{4} \log \frac{p_{AB}(+,+)\,p_{AB}(-,-)\,(1-r)^2}{p_{AB}(+,-)\,p_{AB}(-,+)\,(1+r)^2}. \]
Hence, $f'(r)=0$ if and only if the argument of the logarithm is unity, i.e., if and only if
\[
[(1+x)(1+y) +r]\, [(1-x)(1-y) +r]\,(1-r)^2 = [(1+x)(1-y) -r]\,[(1-x)(1+y) -r]\,(1+r)^2. \]
Expanding and simplifying yields two possible solutions: $r=0$, or $r=(x^2+y^2-x^2y^2)/(2xy)$.  However, in the latter case one has
\[ |r+xy| = \frac{x^2+y^2+x^2y^2}{2|xy|} = \frac{\alpha}{\gamma} +\frac{\gamma}{2} \geq 1+\frac{\gamma}{2} \geq 1, \]
where $\alpha$ and $\gamma$ denote the arithmetic mean and geometric mean, respectively, of $x^2$ and $y^2$ (hence $\alpha\geq\gamma$).   This is clearly inconsistent with the positivity condition (\ref{pos}) (unless $x=y=0$, which trivially saturates  Eq.~(\ref{fr}) for all $r$ as noted above).  The  only remaining solution to $f'(r)=0$ is then $r=0$, implying $f(r)$ has a unique maximum or minimum value at $r=0$.  Finally, it is easily checked that it is a minimum, since
\[ f''(0) = \frac{1}{16} \sum_{a,b} \frac{1}{p_A(a)p_B(b)} - 1 = \frac{1}{16P_A(+)P_A(-)P_B(+)P_B(-)}-1=\frac{1}{(1-x^2)(1-y^2)}-1 \geq 0 \]
(with equality only for the trivially-saturating case $x=y=0$).  Thus, $f(r)\geq f(0)=0$ as required.

\subsection{Application: Resources for Simulating Bell Inequality Violation}

The hallmark feature of quantum correlations is that they cannot be explained by any underlying statistical model that satisfies three physically very plausible properties: (i) no signaling faster than the speed of light, (ii) free choice of measurement settings, and (iii) independence of local outcomes.  Various interpretations of quantum mechanics differ in regard to which of these properties should be given up.  It is of interest to consider by {\it how much} they must be given up, in terms of the information-theoretic resources required to simulate a given quantum correlation.  For example, how many bits of communication, or bits of correlation between the source and the measurement settings, or bits of correlation between the outcomes, are required?  The lower bound for classical mutual information in Eq.~(\ref{c2}) is relevant to the last of these questions.

In more detail, if $P_{AB}(a,b)$ denotes the joint probability of outcomes $a$ and $b$, for measurements of variables $A$ and $B$ on respective spacelike-separated systems, and $\lambda$ denotes any underlying variables relevant to the correlations, then Bayes theorem implies that
\[ P_{AB}(a,b) = \sum_\lambda p_{AB}(\lambda) \, P_{AB}(a,b|\lambda), \]
where summation is replaced by integration over any continuous values of $\lambda$.  The no-signaling property requires that the underlying marginal distribution of $A$, $p_A(a|\lambda)$, is independent of whether $B$ or $B'$ was measured on the second system (and vice versa), while the free-choice property requires that  $\lambda$ is independent of the choice of the measured variables $A$ and $B$, i.e., that
$p_{AB}(\lambda) = p_{A'B'}(\lambda)$
for any $A,A',B,B'$.  Finally, the outcome independence property requires that any observed correlation between $A$ and $B$ arises from ignorance of the underlying variable, i.e., that $P_{AB}(a,b|\lambda) = P_{A}(a|\lambda)\, P_{B}(b|\lambda)$  for all $A$, $B$ and $\lambda$.  Thus the correlation distance of $P_{AB}(a,b|\lambda)$ vanishes identically:
\begin{equation} \label{out}  
C(P_{AB|\lambda})\equiv 0 .
\end{equation}

As is well known, the assumption of all three properties implies that two-valued random variables with values $\pm 1$ must satisfy the Bell inequality \cite{chsh}
\begin{equation} \label{bell}  
\langle AB\rangle + \langle AB'\rangle +\langle A'B\rangle -\langle A'B'\rangle \leq 2, 
\end{equation}
whereas quantum correlations can violate this inequality by as much as a factor of $\sqrt{2}$.
It follows that quantum correlations can only be modeled by relaxing one or more of the above properties, as has recently been reviewed in detail in Ref.~\cite{relaxed}.  

For example, assuming that no-signaling and measurement independence hold (as they do in the standard Copenhagen interpretation of quantum mechanics), and defining $C_{\rm max}$ to be the maximum value of $C(P_{AB|\lambda})$ over all $A$, $B$ and $\lambda$,  it can be shown that Eq.~(\ref{bell}) generalises to the tight bound \cite{relaxed}
\begin{equation}  
\langle AB\rangle + \langle AB'\rangle +\langle A'B\rangle -\langle A'B'\rangle \leq \frac{4}{2-C_{\rm max}} .
\end{equation}
 It follows that to simulate a Bell inequality violation $\langle AB\rangle + \langle AB'\rangle +\langle A'B\rangle -\langle A'B'\rangle=2+V$, for some $V>0$, the observers must share random variables having a correlation distance of at least $C_{\rm max}\geq 2V/(2+V)$.  Hence, using the classical lower bound Eq.~(\ref{c2}) (stated without proof in Ref.~\cite{relaxed}), the observers must share a minimum mutual information of
\begin{equation} 
I_{\rm min} = \log 2 -  H\left(\frac{1+C_{\rm max}}{2},\frac{1-C_{\rm max}}{2}\right) \geq \log 2 - H\left( \frac{2+3V}{4+2V},\frac{2-V}{4+2V} \right)  .
\end{equation}
Note this reduces to zero in the limit of no violation of Bell inequality (\ref{bell}), i.e., when $V=0$, and reaches a maximum of 1 bit of information in the limit of the maximum possible violation, $V=2$.

\section{Quantum Correlation Distance and Qubit Entanglement}

The positivity condition (\ref{pos}) may be used to show that the classical correlation distance between any pair of two-valued random variables is never greater than unity, i.e., that $C(P_{AB})=|r|\leq 1$ \cite{relaxed}.  In contrast, the quantum correlation distance between a pair of qubits can be greater than unity, with upper bound $C(\rho_{AB})\leq 3/2$. More generally, one has 
\begin{equation} \label{cqn} 
C(P_{AB})\leq 2(n-1)/n,~~~~~~~~~~C(\rho_{AB}) \leq 2(n^2-1)/n^2 
\end{equation}
for pairs of $n$-valued random variables and $n$-level quantum systems, with saturation corresponding to maximal correlation and maximal entanglement respectively.  Thus, quantum correlations have a quadratic advantage with respect to correlation distance (this is also the case for mutual information, for which one has $I(P_{AB})\leq \log n$ and $I(\rho_{AB})\leq \log n^2$).

Nonclassical values of the quantum correlation distance are closely related to the quintessential nonclassical feature of quantum mechanics: entanglement. In particular,  $C(\rho_{AB})>1$ is a direct signature of qubit entanglement. Indeed, even correlation distances smaller than unity can imply two qubits are entangled, as per the criterion given in Eq.~(\ref{entang}) and shown below.  An explicit formula for qubit correlation distance in terms of the spin covariance matrix, needed for section 5, is also obtained below.

\subsection{Entanglement Criterion}

Recall that the density operator $\rho_{AB}$ of two qubits may always be written in the Fano form \cite{fano}
\begin{eqnarray} \nonumber 
\rho_{AB} &=& \frac{1}{4} \left[ I\otimes I + u.\sigma \otimes I+I\otimes v.\sigma + \sum_{j,k} \langle \sigma_j\otimes\sigma_k\rangle\, \sigma_j\otimes \sigma_k \right] \\ \label{fano}
&=& \rho_A \otimes \rho_B + \frac{1}{4} \sum_{j,k} T_{jk}\, \sigma_j\otimes\sigma_k .
\end{eqnarray}
Here $I$ is the unit operator; $\{\sigma_j\}$ denotes the set of Pauli spin observables on each qubit Hilbert space; the components of the 3-vectors $u$ and $v$ are the spin expectation values $u_j:=\langle \sigma_j\otimes 1\rangle$ and $v:=\langle 1\otimes \sigma_k\rangle$, for $A$ and $B$ respectively; and $T$ denotes the $3\times3$ spin covariance matrix with coefficients
\[ T_{jk} := \langle \sigma_j\otimes \sigma_k\rangle - \langle \sigma_j\otimes I\rangle\,\langle I\otimes\sigma_k\rangle . \]
It immediately follows from Eq.~(\ref{fano}) that the quantum correlation distance may be expressed in terms of the spin covariance matrix as
\begin{equation} \label{qt}
C(\rho_{AB}) = \frac{1}{4} {\rm tr}\left| \sum_{j,k} T_{jk} \,\sigma_j\otimes\sigma_k \right| .
\end{equation}
This expression will be further simplified in subsection 4.2.	

Now consider the case where $\rho_{AB}$ is a separable state, i.e., of the unentangled form $$\rho_{AB}=\sum_\lambda p(\lambda) \,\tau_A(\lambda)\otimes \omega_B(\lambda),$$ for some probability distribution $p(\lambda)$ and local density operators $\{\tau_A(\lambda)\}$, $\{\omega_B(\lambda)\}$.  
Defining $u_j(\lambda):={\rm tr}[\tau_A(\lambda)\sigma_j]$, $v_k(\lambda):={\rm tr}[\omega_B(\lambda)\sigma_k]$ implies $u=\sum_\lambda p(\lambda)u(\lambda)$ and $v=\sum_\lambda p(\lambda)v(\lambda)$, and substitution into Eq.~(\ref{qt}) then yields
\begin{eqnarray} \nonumber
C(\rho_{AB}) &=& \frac{1}{4} \left\|\sum_\lambda p(\lambda)\,[u(\lambda)-u].\sigma\otimes [v(\lambda)-v].\sigma \right\|_1 \\
\nonumber &\leq& \frac{1}{4} \sum_\lambda p(\lambda) \left\| [u(\lambda)-u].\sigma\right\|_1\, \left\| [v(\lambda)-v].\sigma\right\|_1 \\ 
\nonumber &=& \sum_\lambda p(\lambda) |u(\lambda)-u|\,|v(\lambda) -v| \leq \left[ \sum_\lambda p(\lambda) |u(\lambda)-u|^2\right]^{1/2} \left[ \sum_\lambda p(\lambda) |v(\lambda)-v|^2\right]^{1/2}\\
 &=& \left[ \sum_\lambda p(\lambda) |u(\lambda)|^2-|u|^2\right]^{1/2} \left[ \sum_\lambda p(\lambda) |v(\lambda)|^2-|v|^2\right]^{1/2}
\leq \sqrt{(1-u.u)(1-v.v)} . \label{entanguv}
\end{eqnarray}
Note that second line follows from the properties $\|X+Y\|_1\leq \|X\|_1+\|Y\|_1$ and $\|XY\|_1\leq \|X\|_1\|Y\|_1$ of the trace norm; the third line using  $\|X\|_1={\rm tr}[\sqrt{X^\dagger X}]$  and the Schwarz inequality; and the last line via $|u(\lambda)|, |v(\lambda)|\leq 1$. 

Equation (\ref{entanguv}) holds for all separable qubit states.  Hence, a nonclassical value of the correlation distance, $C(\rho_{AB})>1$, immediately implies that the qubits must be entangled.  More generally, noting that $\rho_A=\frac{1}{2}(I+u.\sigma)$ and $\rho_B=\frac{1}{2}(I+v.\sigma)$, one has ${\rm tr}[\rho_A^2]=(1+u.u)/2$, ${\rm tr}[\rho_B^2]=(1+v.v)/2$, and the stronger entanglement criterion (\ref{entang}) immediately follows from Eq.~(\ref{entanguv}). 

The fact that entanglement is required between two qubits, for $C(\rho_{AB})$ to be greater than the maximum possible value of $C(P_{AB})$ for two-valued classical variables, is a nice distinction between quantum and classical correlation distances. It would be of interest to determine whether this result generalises to $n$-level systems.  This would follow from the validity of Eq.~(\ref{entang})  for arbitrary quantum systems.

\subsection{Explicit Expression for $C(\rho_{AB})$}

To explicitly evaluate $C(\rho_{AB})$ in Eq.~(\ref{qt}), let $T=KDL^T$ denote a singular value decomposition of the spin covariance matrix.  Thus, $K$ and $L$ are real orthogonal matrices and $D={\rm diag}[t_1,t_2,t_3]$, with the singular values $t_1\geq t_2\geq t_3\geq 0$ corresponding to the square roots of the eigenvalues of $TT^T$.  Noting that any $3\times 3$ orthogonal matrix is either a rotation matrix, or the product of a rotation matrix with the parity matrix $-I$, one therefore always has a decomposition of the form $T=\pm KDL^T$ where $K$ and $L$ are now restricted to be rotation matrices. Hence, defining unitary operators $U$ and $V$ corresponding to rotations $K$ and $L$, via  $U\sigma_jU^\dagger=\sum_{j,j'} K_{jj'}\sigma_{j'}$ and $V\sigma_jV^\dagger=\sum_{j,j'} L_{jj'}\sigma_{j'}$, and using the invariance of the trace norm under unitary transformations, the quantum correlation distance in Eq.~(\ref{qt}) can be rewritten as
\[ C(\rho_{AB}) = \frac{1}{4} {\rm tr}\left| \pm \sum_j t_j\, U\sigma_jU^\dagger\otimes V\sigma_jV^\dagger\right| = \frac{1}{4} {\rm tr}\left| \sum_j t_j \,\sigma_j\otimes \sigma_j \right| .\]
  Determining the eigenvalues of the Hermitian operator $\sum_j t_j\, \sigma_j\otimes \sigma_j$ is a straighforward $4\times 4$ matrix calcuation using the standard representation of the Pauli sigma matrices. Summing the absolute values of these eigenvalues then yields the explicit expression
\begin{eqnarray} \nonumber
C(\rho_{AB}) &=& \frac{1}{4}\left[ |t_1+t_2+t_3| + |t_1+t_2-t_3|+|t_1-t_2+t_3| + |-t_1+t_2+t_3|\right]\\
&=& \frac{1}{2} \max \{ t_1+t_2+t_3, 2t_1\} \label{exp}
\end{eqnarray}
for the quantum correlation distance, in terms of the singular values of the spin covariance matrix.  

For example, for the Werner state $\rho_{AB}=p|\psi\rangle\langle \psi|+(1-p)/4\,I\otimes I$, where $|\psi\rangle$ is the singlet state and $-1/3\leq p\leq 1$ \cite{werner}, one has $T=-pI$ and hence that $t_1=t_2=t_3=|p|$.  The corresponding correlation distance is therefore $3|p|/2$, which is greater than the classical maximum of unity for $p>2/3$.  

Equation (\ref{exp})  also allows the qubit entanglement criterion (\ref{entang}) to be directly compared with strongest known criterion based on the spin covariance matrix \cite{zzz}:
\begin{equation}
t_1+t_2+t_3 > 2\sqrt{(1-{\rm tr}[\rho_A^2])\,(1-{\rm tr}[\rho_B^2])} .
\end{equation}
For the above Werner state this criterion is tight, indicating entanglement for $p>1/3$.  Hence, the main interest in weaker entanglement criteria based on quantum correlation distance lies in their direct connection with nonclassical values of the classical correlation distance.

\section{Tight Lower Bound for Quantum Mutual Information} 

Here Eq.~(\ref{q2}) is derived for the case $\rho_A=\rho_B=\frac{1}{2}I$. Evidence is provided for the conjecture that Eq.~(\ref{q2}) in fact holds for all two-qubit states, including a partial generalisation of Eq.~(\ref{q2}) when only one of $\rho_A$ and $\rho_B$ is maximally-mixed.

\subsection{Derivation for Maximally-Mixed $\rho_A$ and $\rho_B$}

The tight lower bound for quantum mutual information in Eq.~(\ref{q2}), for maximally-mixed reduced states, is plotted in Figure 2 below [top solid curve].  Also plotted for comparison are the Pinsker lower bound in Eq.~(\ref{pin2}) [bottom solid curve], and classical lower bound in Eq.~(\ref{c2}) [dashed curve].  The dotted vertical line indicates the value of $C_0\approx0.72654$ in Eq.~(\ref{q2}). It is seen that quantum correlations can violate the classical lower bound for correlation distances falling between $C_0$ and 1.  

\begin{figure}[h]
\caption{Lower bounds for the quantum mutual information between two qubits.}
	\centering
		\includegraphics{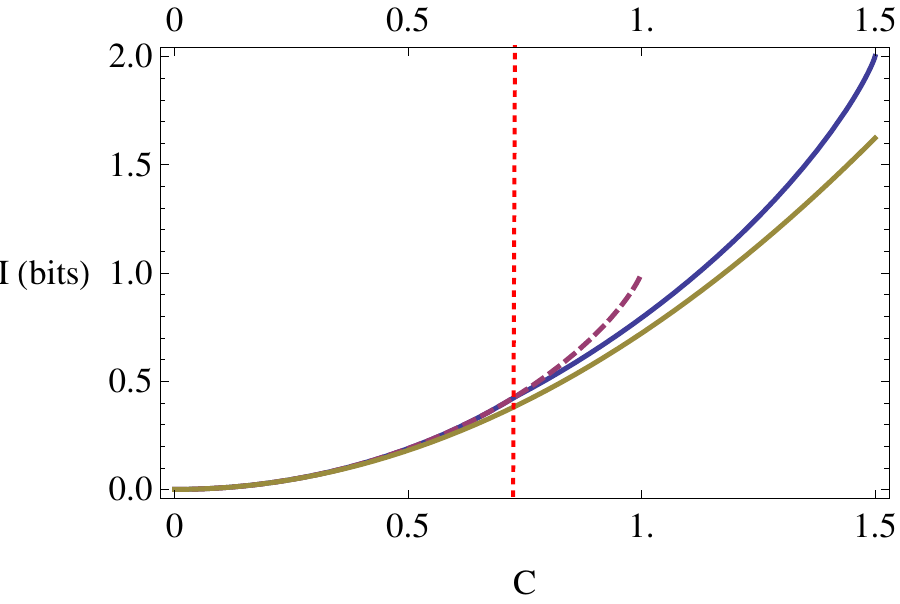}
	\label{fig:pinskerfig2}
\end{figure}

To derive Eq.~(\ref{q2}) for $\rho_A=\rho_B=\frac{1}{2}I$, note first that Eq.~(\ref{fano}) reduces to
$\rho_{AB} = \frac{1}{4}  [ I\otimes I + \sum_{j,k}T_{jk}\, \sigma_j\otimes \sigma_k ]$.
By the same argument given in section 4.2, this can be transformed via local unitary transformations to the state
\begin{equation} \label{rhotil}
\tilde \rho_{AB} = \frac{1}{4}  \left[ I\otimes I + \sum_{j}  r_j\, \sigma_j\otimes \sigma_j\right],
\end{equation}
where $r_j=\alpha t_j$, $\alpha=\pm1$, and $t_1\geq t_2\geq t_3\geq 0$ are the singular values of the spin covariance matrix $T$. Since the quantum mutual information and quantum correlation distance are invariant under local unitary transformations, one has $I(\rho_{AB})=I(\tilde\rho_{AB})$ and  $C(\rho_{AB})=C(\tilde\rho_{AB})$.  Hence Eq.~(\ref{q2}) only needs to be demonstrated for $\tilde\rho_{AB}$.

The mutual information of $\tilde\rho_{AB}$ is easily evaluated as
\begin{equation} \label{itil} 
I(\tilde\rho_{AB}) = S(\tilde\rho_A) + S(\tilde\rho_B) - S(\tilde\rho_{AB}) = \log 4 - H(p_0,p_1,p_2,p_3) ,
\end{equation}
where $p_0 = \frac{1}{4}(1-r_1-r_2-r_3)$,  $p_1=\frac{1}{4}(1-r_1+r_2+r_3)$, $p_2=\frac{1}{4}(1+r_1-r_2+r_3)$, $p_3=\frac{1}{4}(1+r_1+r_2-r_3)$ are the eigenvalues of $\tilde\rho_{AB}$.  Inverting the relation between the $r_j$ and $p_j$ further yields
\begin{equation} \label{cond}
t_j=\alpha r_j=\alpha[1-2(p_0+p_j)],~~~~t_1 +t_2+t_3 = \alpha(1-4p_0) ,
\end{equation}
and hence the correlation distance follows from Eq.~(\ref{exp}) as
\begin{equation} \label{ctil}
C(\tilde\rho_{AB}) = C := \frac{1}{2}\max \{ \alpha(1-4p_0), \alpha(1-4p_0 +1-4p_1)\} . 
\end{equation}

Equation (\ref{itil}) implies that a tight lower bound for $I(\tilde\rho_{AB})$ corresponds to a tight upper bound for $H(p_0,p_1,p_2,p_3)$. To determine the maximum value of $H(p_0,p_1,p_2,p_3)$, for a fixed correlation distance $C$, consider first the case $\alpha=1$.  The ordering and positivity conditions on $t_j$ then require $p_1\leq p_2\leq p_3$, and $p_0+p_j\leq\frac{1}{2}$ for $j=1,2,3$ (implying $p_0\leq 1/4$).  Further, from Eq.~(\ref{ctil}),
$C=\frac{1}{2}\max \{ 1-4p_0, 1-4p_0 + 1-4p_1\}$.
Hence, if $p_1\leq 1/4$, then $C=1-2(p_0+p_1)\leq 1$, implying the constraint $p_0+p_1=(1-C)/2$. Noting the concavity of entropy, the maximum possible entropy under this constraint corresponds to equal values $p_0=p_1=(1-C)/4$, and $p_2=p_3=(1+C)/4$ (which are compatible with the above conditions on the $p_j$).  Conversely, if $p_1\geq 1/4$ then $C=(1-4p_0)/2\leq 1/2$, and hence $p_0=1/4-C/2$ is fixed, implying by concavity that the maximum possible entropy corresponds to $p_1=p_2=p_3=1/4+C/6$ (which again satisfies the required conditions on the $p_j$).  It follows that the maximum possible entropy is (i) the maximum of the entropies $H_1(C)=H((1-C)/4,(1-C)/4,(1+C)/4,(1+C)/4)$ and $H_2(C)=H(1/4-C/2,1/4+C/6,1/4-C/6,1/4-C/6))$ for $C\leq1/2$, and (ii) $H_1(C)$ for $1/2<C\leq 1$.  However, it is straightforward to show  that $H_1(C)\geq H_2(C)$ over their overlapping range. Hence the maximum possible entropy is always $H_1(C)$ for the case $\alpha=1$.

For the case $\alpha=-1$, the conditions on $t_j$ require that $p_1\geq p_2\geq p_3$ and $p_0+p_j\geq\frac{1}{2}$ for $j=1,2,3$ (implying $p_0\geq 1/4$), while from Eq.~(\ref{ctil})
$C=\frac{1}{2}\max \{ 4p_0-1, 4p_0 -1+ 4p_1-1)\}$.  Carrying out a similar analysis to the above, one finds that the maximum possible entropy is (i) the maximum of the entropies $H_1(C)$ and $H_3(C)=H(1/4+Q/2,1/4-Q/6,1/4-Q/6,1/4-Q/6)$ for $C\leq 1$, and (ii) $H_3(C)$ for $1<C\leq 3/2$.  

Numerical comparison shows that $H_3(C)> H_1(C)$ for $C> C_0\approx  0.72654$, and $H_3(C)\leq H_1(C)$ otherwise. Hence, from Eq.~(\ref{itil}) one has the tight lower bound
\begin{equation} \label{q2cop}
 I(\tilde\rho_{AB}) \geq \left\{ \begin{array}{ll}
\log 4 -   H_1(C), &C\leq C_0,\\
\log 4 - H_3(C), &C> C_0.
\end{array} \right. 
\end{equation}
Since $H_1(C)=\log 2+H((1-C)/2,(1+C)/2)$, it follows that Eq.~(\ref{q2})  holds for $\tilde\rho_{AB}$ in Eq.~(\ref{rhotil}), and hence for all qubit states with maximally-mixed reduced density operators, as claimed.

The states saturating the lower bound in Eqs.~(\ref{q2}) and (\ref{q2cop}) are easily constructed from the above derivation.  In particular, they are given by
\begin{equation} \label{sat}
\rho(C) := \left\{ \begin{array}{ll}
\frac{1}{4}\left[ I\otimes I + C\, \sigma_1\otimes \sigma_1\right], &C\leq C_0,\\
\frac{1}{4}\left[ I\otimes I -(2C/3)\, \sum_j \sigma_j\otimes \sigma_j\right], &C> C_0,
\end{array} \right.
\end{equation}
and any local unitary transformations thereof, where the quantum correlation distance of $\rho(C)$ is $C$ by construction.  

Note that $\rho(C)$ is unentangled for $C\leq C_0$ (it can be written as a mixture of $(1/4)I\otimes I$, $|+\rangle\langle+|\otimes |+\rangle\langle+|$ and $|-\rangle\langle-|\otimes |-\rangle\langle-|$, where $\sigma_1|\pm\rangle=\pm|\pm\rangle$). Conversely, $\rho(C)$ is an entangled Werner state for $C\geq C_0$ (with singlet state weighting $p=2C/3>1/3$).  Hence, the lower bound in Eqs.~(\ref{q2}) and (\ref{q2cop}) can only be achieved by entangled states for $C\geq C_0$, and cannot be achieved by any two-valued classical random variables.

\subsection{Conjecture}

It is conjectured that Eq.~(\ref{q2}) is in fact a tight lower bound for any two-qubit state.  This conjecture would follow immediately if it could be shown that
\begin{equation} \label{conj}
I(\rho_{AB}) \geq I(\rho'_{AB})
\end{equation}
for arbitrary $\rho_{AB}$, where $\rho'_{AB}:=\rho_{AB} -\rho_A\otimes\rho_B+(1/4)I\otimes I$. This is because $\rho'_{AB}$ is of the form of $\tilde\rho_{AB}$ in Eq.~(\ref{rhotil}), and hence $I(\rho'_{AB})$ satisfies Eq.~(\ref{q2cop}).

Partial support for Eq.~(\ref{conj}), and hence for the conjecture, is given by noting that any $\rho_{AB}$ and corresponding $\rho'_{AB}$ can be brought to the respective forms
\[ \rho_{AB} =  \rho_A \otimes \rho_B + \frac{1}{4}\sum_j r_j \,\sigma_j\otimes \sigma_j, ~~~~\rho'_{AB}= \frac{1}{4}\left[ I\otimes I + \sum_j r_j \,\sigma_j\otimes \sigma_j\right] \]
via suitable local unitary transformations, similarly to the argument in section 4.2.  Defining the function 
\[ F(r_1,r_2,r_3):= I(\rho_{AB}) - I(\rho'_{AB}), \]
it is straightforward to show that $F=0$ and $\partial F/\partial r_j=0$ for $r_1=r_2=r_3=0$, consistent with $F\geq 0$.  However, it remains to be shown that the gradient $\partial F/\partial r_j=0$ does not vanish for other physically possible values of the $r_j$ (other than for the trivially saturating case $\rho_A=\rho_B=(1/2)I$).

The above conjecture is further supported by the generalisation of Eq.~(\ref{q2}) in the following section.

\subsection{Generalisation to Maximally-Mixed $\rho_A$ or $\rho_B$}

It is straighforward to show that the lower bound on quantum mutual information is tight for $C\geq C_0$ when just {\em one} of the mixed density operators is mixed, i.e., if  $\rho_A$ {\rm or} $\rho_B$ is equal to $(1/2)I$. 

First, since $(1/2)I$ is invariant under unitary transformations, the same argument as in section 4.2 implies the state can always be transformed by local unitary transformations to the generalised form
\[ \tilde\rho_{AB} = \frac{1}{4} \left[ \tilde\rho_A \otimes \tilde\rho_B + \alpha\sum_j t_j\,\sigma_j\otimes \sigma_j\right] \]
of Eq.~(\ref{rhotil}), where either $\tilde\rho_A$ or $\tilde\rho_B$ equals $(1/2)I$ and $\alpha=\pm 1$.  

Second, let ${\cal T}$ denote the `twirling' operation, corresponding to applying a random unitary transformation of the form $U\otimes U$ \cite{twirl}.  It is easy to check that by definition ${\cal T}(I\otimes I)=I\otimes I$, ${\cal T}(I\otimes \sigma_j)=0={\cal T}(\sigma_j\otimes 1)$ and ${\cal T}(\sigma_j\otimes\sigma_j)={\cal T}(\sigma_k\otimes\sigma_k)$, for any $j$ and $k$.  Since Werner states are invariant under twirling \cite{werner,twirl}, it follows that ${\cal T}(\sigma_j\otimes\sigma_j) = (1/3)\sum_k\sigma_k \otimes \sigma_k$. Using these properties, one finds that ${\cal T}(\tilde\rho_A\otimes\tilde\rho_B)=(1/4)I\otimes I$ if one of $\tilde\rho_A$ or $\tilde\rho_B$ is maximally mixed, and hence that
\[ {\cal T}(\tilde{\rho}_{AB}) =  \frac{1}{4} \left[ I\otimes I + \alpha \bar{t} \,\sum_j\sigma_j\otimes \sigma_j \right]= \rho(-3\alpha \bar{t}/2),\]
where $\bar{t}:=(t_1+t_2+t_3)/3$ and the second equality holds for $C\geq C_0$ (but not otherwise), with $\rho(C)$ defined as per Eq.~(\ref{sat}).  
Further, from Eq.~(\ref{exp}) one has 
\[ C({\cal T}(\tilde{\rho}_{AB})) = C = \frac{1}{2}\max\{ 2\bar{t},3\bar{t}\} = 3\bar{t}/2. \]
Recalling that $\rho(C)$ saturates Eq.~(\ref{q2cop}), an analysis similar to section 5.1 shows for $C\geq C_0$ that
\[ I({\cal T}(\tilde{\rho}_{AB})) = \log 4 - H_3(-\alpha C) \geq \log 4 - H_3(C), \]
with equality for $\alpha=-1$.

Third, again using  ${\cal T}(\tilde\rho_A\otimes\tilde\rho_B)=(1/4)I\otimes I$, and the property that the relative entropy is non-increasing under the twirling operation, it follows that
\begin{equation} 
I(\tilde\rho_{AB}) = S(\tilde\rho_{AB}\| \tilde\rho_A\otimes\tilde\rho_B)\geq S({\cal T}(\tilde\rho_{AB})\|{\cal T}( \tilde\rho_A\otimes\tilde\rho_B))=I({\cal T}(\tilde\rho_{AB})) \geq \log 4 - H_3(C) 
\end{equation}
for $C\geq C_0$.  Since Werner states are invariant under twirling, this inequality is tight for $\alpha=-1$, being saturated by the choice $\tilde\rho_{AB}=\rho(C)$.
Recalling that mutual information and correlation distance are invariant under local unitary operations, the inequality is therefore tight for any $\rho_{AB}$ for which one of $\rho_A$ and $\rho_B$ is maximally mixed, as claimed.

\section{Classically-Correlated Quantum States}

It is well known that a quantum system behaves classically if the state and the observables of interest all commute, i.e., if they can be simultaneously diagonalised in some basis.  Hence, a joint state will behave classically if the relevant observables of each system commute with each other and the state.  It is therefore natural to define $\rho_{AB}$ to be {\it classically correlated} if and only if it can be diagonalised in a joint basis \cite{horo}, i.e., if and only if
\begin{equation} \label{class}
\rho_{AB} = \sum_{j,k} P(j,k) |j\rangle\langle j|\otimes |k\rangle\langle k|
\end{equation}
for some distribution $P(j,k)$ and orthonormal basis set $\{|j\rangle\otimes|k\rangle\}$.  Classical correlation is preserved by tensor products, and by mixtures of commuting states.

While, strictly speaking, a classically-correlated quantum state only behaves classically with respect to observables that are diagonal with respect to $|j\rangle\otimes |k\rangle$, they also have a number of classical correlation properties with respect to general observables \cite{horo,molmer}, briefly noted here.

First, $\rho_{AB}$ above is separable by construction, and hence is unentangled. Second, since it is diagonal in the basis $\{|j\rangle\otimes|k\rangle\}$, the mutual information and correlation distance are easily calculated as
\begin{equation}
I(\rho_{AB})=I(P),~~~~~~C(\rho_{AB}) = C(P),
\end{equation}
and hence can only take classical values.  

Third, if $M$ and $N$ denote any observables for systems $A$ and $B$ respectively, then their joint statistics are given by
\[ P_{MN}(m,n) = \sum_{j,k} p(m|j)\, p(n|k)\, P(j,k) = \sum_{j,k} {\cal S}_{m,n;j,k}\, P(j,k), \]
where ${\cal S}_{m,n;j,k}=p(m|j)\,p(n|k)$ is a stochastic matrix with respect to its first and second pairs of indices. Similarly, one finds
\[ P_M(m)\,P_N(n) = \sum_{j,k} {\cal S}_{m,n;j,k}\, P(j)\,P(k) \]
for the product of the marginals.  Since the classical relative entropy and variational distance can only decrease under the action of a stochastic matrix, it  follows that one has the tight inequalities \cite{horo,molmer}
\begin{equation}
 I(P_{MN}) \leq I(P) =I(\rho_{AB}),~~~~~C(P_{MN})\leq C(P) =C(\rho_{AB}) ,
\end{equation}
with saturation for $M$ and $N$ diagonal in the bases $\{|j\rangle\}$ and $\{|k\rangle\}$ respectively.  Maximising the first of these equalities over $M$ or $N$ immediately implies that classically-correlated states have zero quantum discord.

Finally, for two-qubit systems, Eq.~(\ref{class}) implies that $\rho_{AB}$ is classically correlated if and only if it is equivalent under local unitary transformations to a state of the form
\[ \rho'_{AB} = \frac{1}{4}\left[(1+x\sigma_1)\otimes(1+y\sigma_1) +r\,\sigma_1\otimes\sigma_1\right], \]
where $x,y\in[-1,1]$ and $r$ satisfies Eq.~(\ref{pos}).  Hence, the mutual information is bounded by the classical lower bound in Eq.~(\ref{c2}), and $\rho(C)$ in Eq.~(\ref{sat}) is classically correlated for $C\leq C_0$.  It follows that the lower bound for quantum mutual information in Eq.~(\ref{q2}) can be attained by classically-correlated states if $C\leq C_0$.  Conversely, the minimum possible bound cannot be reached by any classically-correlated two-qubit state if $C>C_0$.

\section{Conclusion}

Lower bounds for mutual information have been obtained that are stronger than those obtainable from general bounds for relative entropy and variational distance. Unlike the Pinsker inequality in Eq.~(\ref{pin2}), the quantum form of these bounds is not a simple generalisation of the classical form.  

Similarly to the case of upper bounds for (classical) mutual information \cite{upper}, the tight lower bounds obtained here depend on the dimension of the systems.  The results of this paper represent a preliminary investigation largely confined to  two-valued classical variables and qubits.  It would be of interest to generalise both the classical and quantum cases, and to further investigate connections between them.

Open questions include whether a quantum correlation distance greater than the corresponding maximum classical correlation distance is a signature of entanglement for higher-dimensional systems, and whether the related qubit entanglement criterion in Eq.~(\ref{entang}) holds more generally. The conjecture in section 5.2, as to whether the quantum lower bound in Eq.~(\ref{q2}) is valid for all two-qubit states, also remains to be settled.  Finally, it would be of interest to generalise and to better understand the role of the transition from classically-correlated states to entangled states in saturating information bounds, in the light of Eq.~(\ref{sat}) for qubits.
\\
~
\\
{\bf Acknowledgements}:
This research was supported by 
the ARC Centre of Excellence CE110001027.

\bibliographystyle{mdpi}
\makeatletter
\renewcommand\@biblabel[1]{#1. }
\makeatother


\end{document}